\documentclass[a4paper, amsfonts, amssymb, amsmath, reprint, showkeys, nofootinbib, twoside]{revtex4-2}
\usepackage[english]{babel}
\usepackage[utf8]{inputenc}
\usepackage[colorinlistoftodos, color=green!40, prependcaption]{todonotes}
\usepackage{amsthm}
\usepackage{mathtools}
\usepackage{physics}
\usepackage{xcolor}
\usepackage{graphicx}
\usepackage[left=23mm,right=13mm,top=35mm,columnsep=15pt]{geometry} 
\usepackage{adjustbox}
\usepackage{placeins}
\usepackage[T1]{fontenc}
\usepackage{lipsum}
\usepackage{csquotes}
\usepackage[pdftex, pdftitle={Article}, pdfauthor={Author}]{hyperref} 
\newcommand{\aap}{Astron. Astrophys.}
\newcommand{\mnras}{Mon. Not. R. Astron. Soc.}
\newcommand{\pasj}{Publ. Astron. Soc. Jpn.}
\newcommand{\jcap}{J. Cosmology Astropart. Phys.}
\newcommand{\physrep}{Phys. Rep.}

\newcommand{\beq}{\begin{equation}}
\newcommand{\eeq}{\end{equation}}

\newcommand{\bfx}{\mbox{\boldmath{$x$}}}

\newcommand{\bfk}{\mbox{\boldmath{$k$}}}

\newcommand{\Mpc}{\mathrm{Mpc}}

\newcommand{\RT}[1]{#1}
\newcommand{\rev}[1]{#1}  

\begin{document}
\title{Cosmological boost factor for dark matter annihilation at redshifts of $z=10$--$100$ using the power spectrum approach}


\author{Ryuichi Takahashi$^1$}
\author{Kazunori Kohri$^{2,3,4}$}

\affiliation{$^1$Faculty of Science and Technology, Hirosaki University, 3 Bunkyo-cho, Hirosaki, Aomori 036-8561, Japan \email{takahasi@hirosaki-u.ac.jp}}

\affiliation{$^2$Institute of Particle and Nuclear Studies, KEK, 1-1 Oho, Tsukuba, Ibaraki 305-0801, Japan}
\affiliation{$^3$The Graduate University for Advanced Studies (SOKENDAI), 1-1 Oho, Tsukuba, Ibaraki 305-0801, Japan}
\affiliation{$^4$Kavli IPMU (WPI), UTIAS, The University of Tokyo,
Kashiwa, Chiba 277-8583, Japan}

\date{\today} 

\begin{abstract}
We compute the cosmological boost factor at high redshifts of $z=10$--$100$ by integrating the non-linear matter power spectrum measured from high-resolution cosmological $N$-body simulations. 
An accurate boost factor is required to estimate the energy injection from dark matter annihilation, which may affect the cosmological re-ionization process.
We combined various box-size simulations (side lengths of $1 \, {\rm kpc}$--$10 \, {\rm Mpc}$) to cover a wide range of scales, i.e. $k=1$--$10^7 \, {\rm Mpc}^{-1}$.
The boost factor is consistent with the linear theory prediction at $z \gtrsim 50$ but strongly enhanced at $z \lesssim 40$ as a result of non-linear matter clustering.
Although dark matter free-streaming damping was imposed at $k_{\rm fs}=10^6 \, {\rm Mpc}^{-1}$ in the initial power spectrum, the damping disappears at later times of $z\lesssim40$ as a result of the power transfer from large to small scales. 
Because the simulations do not explore very small-scale clustering at $k>10^7 \, \Mpc^{-1}$, our result is a \textit{lower bound} on the boost factor at $z \lesssim 40$.  
A simple fitting function of the boost factor is also presented.
\end{abstract}


\maketitle

\section{Introduction}

\RT{In the standard cold dark matter model, cosmological structure formation is driven by the gravitational force of dark matter.}
The nature of dark matter remains elusive but may comprise unknown elementary particles (e.g. \cite{BH2018}).
The annihilation of dark-matter particles may generate high-energy photons or particles (such as $e^+ e^-$ and $q \bar{q}$). 
\RT{\rev{At the photon} decoupling epoch (at $z \approx 1000$), the dark matter density was nearly homogeneous; however, at later times, the density contrast evolved and high density regions (such as halos) formed, where the annihilation is enhanced.
The high-energy photon production and intergalactic medium heating from} the annihilation at high redshifts of $z \gtrsim 6$ could affect the cosmological re-ionization process~\cite{Valdes2013,Evoli2014,Poulin2015,Liu2016,Short2020}.
Recently, the EDGES experiment reported the first detection of an absorption signature in radio signals at $z=17$~\cite{Bowman2018}, which indicates a lower gas temperature than the background radiation.
This measurement can constrain (or exclude) the energy injection from dark matter annihilation~\cite{DAmico2018,Yang2018,Cheung2019,Sekiguchi2021}.

The annihilation rate is proportional to the square of the dark matter density; therefore, an inhomogeneous density will enhance dark matter annihilation.
Let us denote the dark-matter density at a comoving coordinate $\bfx$ and a redshift $z$ as $\rho(\bfx;z)$.
This density can be decomposed into its spatial mean $\bar{\rho}(z)$ and its density contrast $\delta (\bfx;z)$ such that $\rho(\bfx;z)=\bar{\rho}(z) [1+\delta (\bfx;z)]$. 
Because the collision rate 
is proportional to $\rho^2$, its spatial average over the universe is
\beq
 \langle \rho^2(\bfx;z) \rangle 
 = \bar{\rho}^2(z) B(z),
 \label{BF_def}
\eeq
where the cosmological boost factor is defined as $B(z) \equiv 1+ \langle \delta^2(\bfx;z) \rangle$.

Two methods have been used to calculate $B(z)$: the halo-model approach (e.g., \cite{Ulino2002,TS2003,Ando2006,Cirelli2011,Shirasaki2014,Hutten2018}) and the power spectrum (PS) approach~\cite{Serpico2012,Sefusatti2014}.
Both approaches give consistent results (e.g. Fig.~1 in Ref.~\cite{Fermi2015}).
In the former approach, the boost factor for a single halo is calculated and all the contributions from multiple halos for the given model parameters, such as the halo density profile and the mass function (e.g.~\cite{CS2002}), are summed.
However, there are several known uncertainties, including the ellipticity of the halo shape, inner density profile, halo mass function, subhalo (and sub-subhalo) abundance, and baryonic feedback effects.
In fact, substructure clumps enhance dark matter annihilation~\cite{Berezinsky2003,Diemand2007,Springel2008,Zavala2016,Hiroshima2018,Ando2019} and gas cooling increases the central density of the halos~\cite{Vogel2014,Schaller2016,Chua2019}.
This makes theoretical modelling very complicated.
Furthermore, the model predictions (such as the mass function and density profile) must be extrapolated to very small scales that cannot be resolved by current (or even near-future) numerical simulations. 
These model uncertainties cause orders of magnitude variations in $B(z)$ (e.g.~\cite{Mack2014,Sanchez2014,Moline2017}). 

In the latter approach, $B(z)$ is obtained by integrating the matter PS with respect to the wavenumber of the density fluctuations.
This approach, first proposed by Refs.~\cite{Serpico2012,Sefusatti2014}, is much simpler and has fewer uncertainties (such as very small scale clustering and baryonic effects) than the former approach. 
References~\cite{Serpico2012,Sefusatti2014} calculated $B(z)$ to estimate extra-galactic gamma-ray flux from dark matter annihilation at $z=0$--$6$.
They prepared the non-linear PS using several methods: a fitting formula (Halofit~\cite{Smith2003,Takahashi2012}), the stable clustering ansatz~\cite{Peebles1980} and Millennium Simulations I and II~\cite{Springel2005,Boylan2009}.
They extrapolated the analytical PS to very small scales and then integrated the PS up to the free-streaming scale of dark matter ($\sim 10^7 \, {\rm Mpc}^{-1}$).

In this paper, we calculate $B(z)$ at redshifts of $z=10$--$100$ using the matter PS measured from high-resolution cosmological $N$-body simulations.
We run different box size simulations (with cubic-box side lengths of $L=1 \, {\rm kpc}$, $10 \, {\rm kpc}$, $100 \, {\rm kpc}$, $1 \, {\rm Mpc}$, and $10 \, {\rm Mpc}$) to cover a wide range of scales up to $k=1.6 \times 10^7 \, \Mpc^{-1}$.
These are dark matter only simulations; however, the baryonic effect is included in the initial PS. 
The simulations follow non-linear evolution near the free-streaming scale, which is set to $k_{\rm fs}=10^6 \, \Mpc^{-1}$.
Therefore, our analysis does not rely on extrapolation beyond the free-streaming scale.

There has been several studies of first halo formation near the free-streaming scale using $N$-body simulations~\cite{Diemand2005,Ishiyama2014,Schneider2015}.   
These studies indicate that Earth-mass halos with $\approx 10^{-6} M_\odot \, [k_{\rm fs}/(10^6 \, \Mpc^{-1})]^{-3}$ form at $z \approx 30$.
\rev{Recently, Ref.~\cite{Wang2020} performed multi-scale zoom-in simulations at $z=0$ covering the halo-mass range from $10^{-6} \, M_\odot$ to $10^{15} \, M_\odot$.}
The primary interest of these studies was the halo properties, such as the mass function and density profile.
As far as we know, no one has studied the non-linear evolution of the PS near the free-streaming scale. 


The rest of this paper is organized as follows. Section II discusses the cosmological boost factor in the PS approach and our simulation setting. 
Section III presents our main results: the non-linear matter PS measured from the simulations and the resulting boost factor.
Section IV discusses the effects of density fluctuations larger than the simulation box, the small-scale PS in the halo model, and baryonic effects on PS. 
Section V summarizes our study.

Throughout this paper, we adopt a cosmological model consistent with the \textit{Planck} 2015 best-fit flat $\Lambda$CDM model~\citep{Planck2015}: a matter density of $\Omega_{\rm m}=1-\Omega_{\Lambda}=0.3089$, a baryon density of $\Omega_{\rm b}=0.0486$, a Hubble parameter of $h=0.6774$, a spectral index of $n_{\rm s}=0.9667$, and an amplitude of matter density fluctuations on the scale of $8 \, h^{-1} \, \Mpc$ $\sigma_8=0.8159$.

\section{Cosmological boost factor}

This section introduces the PS approach (Subsection II.A) and then discusses the linear PS (subsection II.B) and our $N$-body simulation settings (Subsection II.C).

\subsection{PS approach}

Let us denote the Fourier transform of the dark matter density fluctuations as $\tilde{\delta}(\bfk;z)$, where $\bfk$ is the wavevector in the comoving scale.
Then, the matter PS is defined as $\langle \tilde{\delta}(\bfk;z) \tilde{\delta}(\bfk^\prime;z) \rangle \equiv (2 \pi)^3 P(k;z) \, \delta_{\rm D}(\bfk+\bfk^\prime)$, where $\delta_{\rm D}$ is the Dirac delta function.
The dimensionless matter PS is defined as $\Delta^2(k;z) \equiv k^3 P(k;z)/(2 \pi^2)$.   
Then, using the Fourier transform, 
the cosmological boost factor at a redshift $z$ is~\cite{Serpico2012,Sefusatti2014}
\beq
  B(z) = 1+\int_0^\infty \! d\ln \! k \, \Delta^2(k;z).
\label{boost_fac}
\eeq
In the linear theory, because $\Delta_{\rm L}^2(k;z) \propto k^{n_{\rm s}+3}$ in the low-$k$ limit and $\Delta_{\rm L}^2(k;z) \rightarrow 0$ in the high-$k$ limit (over the free-streaming scale), 
the integral in Eq.~(\ref{boost_fac}) converges.
However, in the non-linear regime, $N$-body simulations are required to obtain $\Delta^2(k;z)$ in the high-$k$ regime; this is discussed in the following sections.

\subsection{Linear PS}

\begin{figure}
\centering\includegraphics[width=9.5cm]{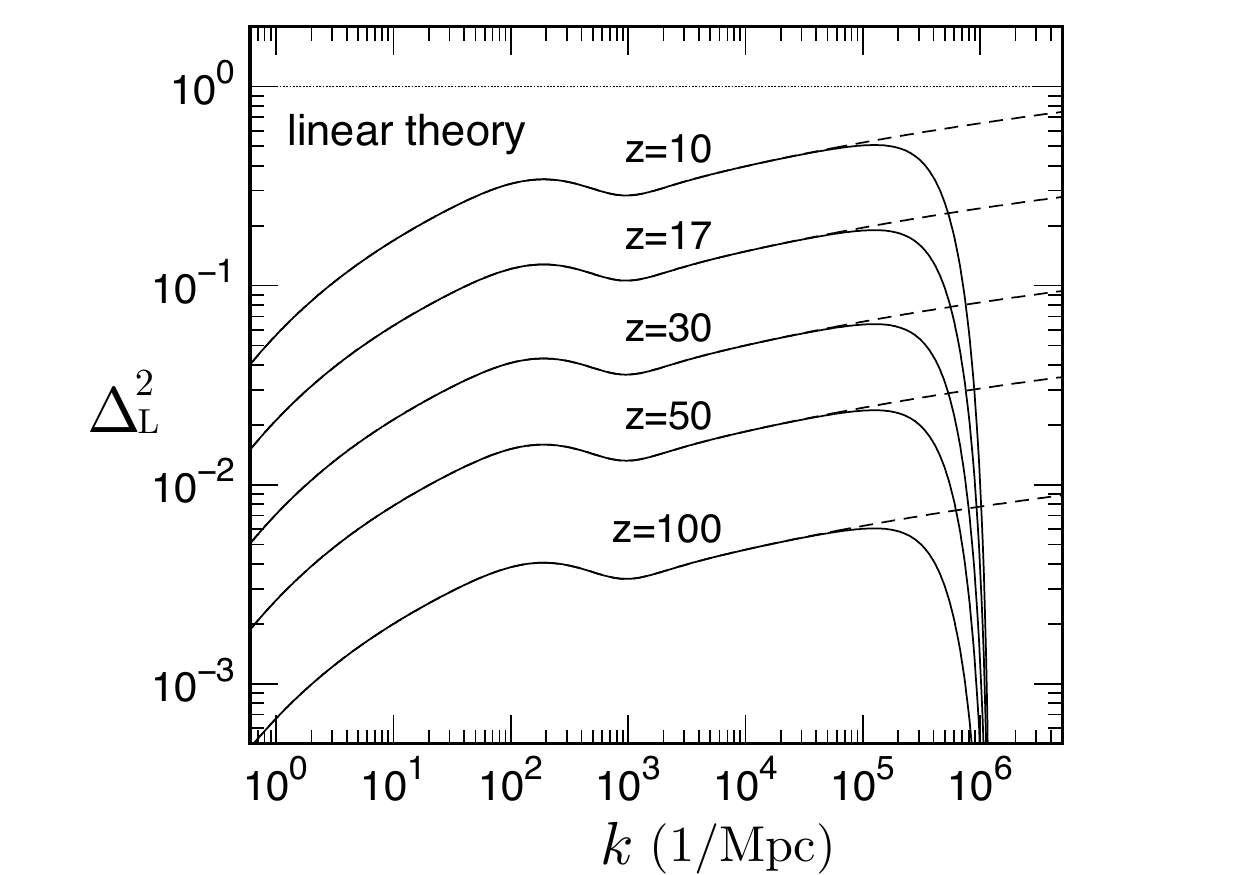}
\caption{Dimensionless linear matter power spectrum at $z=10$--$100$. 
Solid curves include free-streaming damping at $k_{\rm fs}=10^6 \, {\rm Mpc}^{-1}$, whereas dashed curves exclude free-streaming damping.
The suppression at $k \gtrsim 10^3 \, \Mpc^{-1}$ is caused by the baryon gas pressure (i.e. the Jeans effect) after the decoupling epoch~\cite{Yamamoto1998}.}
\label{fig_delk_linear}
\end{figure}

The linear matter PS is written as
\beq
  P_{\rm L}(k;z)=A  \, \left( \frac{k}{k_\ast} \right)^{n_{\rm s}} D^2_+(z) \, T^2(k;z) \, D_{\rm fs}^2(k),
\label{linear_pk}
\eeq
where $k_\ast=0.05 \, {\rm Mpc}^{-1}$ is the pivot scale and $D_+(z)$ is the linear growth factor, which is safely approximated as $D_+(z) \propto (1+z)^{-1}$ in the redshift range of $z=10$--$100$.
The amplitude $A$ and the spectral index $n_{\rm s}$ are set to be consistent with the {\it Planck} 2015 result~\cite{Planck2015}.
Here, we do not consider the running (or the running of running) of the spectral index.

$T(k;z)$ is the transfer function for the total matter density (i.e. dark matter and baryons).
Here, we use the fitting function $T(k;z)$ given in Appendix C of Ref.~\cite{Yamamoto1998}, obtained from cosmological perturbation theory\footnote{There is a minor typo in their formula (Kazuhiro Yamamoto, private communication). In their Appendix C, $\alpha_1$ should be replaced with $\alpha_1=[1-(1+24 \, \Omega_{\rm c}/\Omega_{\rm m})^{1/2}]/4$, where $\Omega_{\rm c}=\Omega_{\rm m}-\Omega_{\rm b}$.}.
In their study, the Boltzmann equation was numerically solved and the results were fitted down to a very small scale ($k = 10^4 \, {\rm Mpc}^{-1}$). 
The quoted accuracy of the fitting formula is a few percent ($10 \, \%$) for $k=1$--$100 \, {\rm Mpc}^{-1}$ ($k>100 \, {\rm Mpc}^{-1}$).
We confirmed that their $T(k;z)$ agrees with the {\tt CAMB} output~\cite{Lewis2000} within $8 \, \%$ for $k<10^4 \, \Mpc^{-1}$. 
At small $k \, (\lesssim 0.1 \, {\rm Mpc}^{-1})$, the $T(k;z)$ is consistent with the Bardeen-Bond-Kaiser-Szalay formula~\cite{BBKS1986} with the baryonic correction~\cite{HS1996}. 
After the decoupling epoch, the baryonic gas pressure suppresses the growth of density fluctuations smaller than the Jeans length.
As time continues, the gas temperature (and pressure) decreases, and thus, the Jeans length decreases. 
Therefore, the transfer function depends on the redshift (i.e. the suppression is more significant for lower $z$; see also Fig.~5 in Ref.~\cite{Yamamoto1998}).
Because we are interested in the matter clustering at $z\approx10$, we used the transfer function at $z=10$, $T(k;z=10)$, throughout this paper.
In this case, $P_{\rm L}(k;z)$ simply evolves in proportion to $D_+^2(z)$.

The damping factor due to the dark matter free streaming, $D_{\rm fs}^2(k)$, is taken from \cite{Green2004}:
\beq
 D_{\rm fs}(k) = \left[ \, 1-\frac{2}{3} \left( \frac{k}{k_{\rm fs}} \right)^2 \right] \exp \left[ -\left( \frac{k}{k_{\rm fs}} \right)^2 \right],
\eeq
for $k<\sqrt{3/2} \, k_{\rm fs}$, and $D_{\rm fs}(k)=0$ otherwise.
Throughout this paper, the free-streaming scale is set to $k_{\rm fs}=10^6 \, {\rm Mpc}^{-1}$, \RT{which corresponds to a kinetic decoupling temperature of $T_{\rm d} \approx 1 \, {\rm MeV}$ for dark matter particles (e.g.~\cite{Green2004,Loeb2005}).} 
Note that the primordial PS is currently measured up to $k \simeq 0.2 \, {\rm Mpc}^{-1}$ by \textit{Planck} \cite{Planck2018Inflation}, and therefore the linear PS in Eq. (\ref{linear_pk}) relies on an extrapolation up to $k_{\rm fs}$. 

The dimensionless linear PS, $\Delta_{\rm L}^2(k;z)$, is plotted in Fig.~\ref{fig_delk_linear}. 
The suppression at $k \gtrsim 10^3 \, {\rm Mpc}^{-1}$ is caused by the Jeans effect after the decoupling epoch (see also Fig.~4 in Ref.~\cite{Yamamoto1998}). 
This scale is determined by the Jeans length $\lambda_{\rm J}$ just after the decoupling epoch: $\lambda_{\rm J}=2\pi/k_{\rm J}$, where $k_{\rm J} = 9 \times 10^2 \, (\Omega_{\rm m} h^2)^{1/2} \, \Mpc^{-1}$~\cite{Yamamoto1998}.
Even at $z=10$, the amplitude of $\Delta_{\rm L}$ is less than unity over the entire $k$ range.
For the quasi-nonlinear regime ($\Delta_{\rm L} \gtrsim 0.1$), $N$-body simulations are required to follow the non-linear evolution. 

\subsection{$N$-body simulations}

\begin{table*}
\caption{Summary of our $N$-body simulations: the side length of cubic simulation box $L$, the number of particles $N_{\rm p}$, the minimum wavenumber $2 \pi/L$, the particle Nyquist wavenumber \rev{$k_{\rm Ny} \equiv (\pi/L) N_{\rm p}^{1/3}$}, and the $N$-body particle mass $m_{\rm p}$. Values in parentheses indicate differing values for the low-resolution runs.}
\label{table_Nbody}
\centering
\begin{tabular}{ccccc}
\hline
 $L$ & $N_{\rm p}$ & $2 \pi/L$ [${\rm Mpc}^{-1}$] & $k_{\rm Ny}$ [${\rm Mpc}^{-1}$] & $m_{\rm p} \, [M_\odot]$ \\
\hline
$10 \, {\rm Mpc}$ & ~$5120^3$ ($2560^3$) & $0.63$ & $1.6 \!\times\! 10^3$ ($800$)  & ~$29$ \, ($230$) \\
$1 \, {\rm Mpc}$ & ~$5120^3$ ($2560^3$) & $6.3$ & $1.6 \!\times\! 10^4$ ($8.0 \!\times\! 10^3$)  & ~$2.9 \!\times\! 10^{-2}$ ($0.23$) \\
$100 \, {\rm kpc}$ & ~$5120^3$ ($2560^3$) & $63$ & $1.6 \!\times\! 10^5$ ($8.0 \!\times\! 10^4$)  & ~$2.9 \!\times\! 10^{-5}$ ($2.3 \!\times\! 10^{-4}$) \\
$10 \, {\rm kpc}$ & ~$5120^3$ ($2560^3$) & $630$ & $1.6 \!\times\! 10^6$ ($8.0 \!\times\! 10^5$)  & ~$2.9 \!\times\! 10^{-8}$ ($2.3 \!\times\! 10^{-7}$) \\
$1 \, {\rm kpc}$ & ~$5120^3$ ($2560^3$) & $6.3 \!\times\! 10^3$ & $1.6 \!\times\! 10^7$ ($8.0 \times 10^6$) & ~$2.9 \!\times\! 10^{-11}$ ($2.3 \!\times\! 10^{-10}$)  \\
\hline
\end{tabular}
\end{table*}

To obtain the non-linear $\Delta^2$, we ran $N$-body simulations in cubic boxes to follow the gravitational evolution of collisionless particles.
These are dark matter only simulations \RT{(i.e. without non-linear baryonic processes such as star formation, gas cooling, or radiative transfer)}.
However, the baryonic effects in the initial linear PS \RT{(such as the baryon acoustic oscillation and the Silk damping)} are included.
\RT{Baryonic effects on the non-linear PS are discussed in Subsection IV.C.}
Because the length scales of interest are broad, $k = 1$--$10^7 \, {\rm Mpc}^{-1}$, we combined five different box size simulations with side lengths of $L=1 \, {\rm kpc}, 10 \, {\rm kpc}, 100 \, {\rm kpc}, 1 \, {\rm Mpc},$ and $10 \, {\rm Mpc}$. 
The number of particles in each box was $N_{\rm p}=5120^3$ and $2560^3$ for the high-resolution (HR) and low-resolution (LR) runs, respectively.
These different resolution runs were used to check the numerical convergence, given the finite spatial resolution.
The minimum wavenumber was $k_{\rm min}=2 \pi/L$, which is necessarily smaller than $k_{\rm fs}$ to include the initial power at $k<k_{\rm fs}$.
The simulation results are reliable up to the particle Nyquist wavenumber, given by \rev{$k_{\rm Ny} \equiv (\pi/L) N_{\rm p}^{1/3}$}. 
Our simulation settings, including the values of $L$, $N_{\rm p}$, $k_{\rm min}$, $k_{\rm Ny}$, and the $N$-body particle mass $m_{\rm p}$, are summarized in Table \ref{table_Nbody}.
\RT{The particle mass in the smallest box ($\simeq 3 \times 10^{-11} \, M_\odot$) is small enough to resolve the minimum halo mass determined by $k_{\rm fs}$ (i.e. Earth mass $\sim 10^{-6} \, M_\odot$).}

The initial particle positions were \rev{given with the grid-based configuration} 
on the basis of the second-order Lagrangian perturbation theory~\cite{Crocce2006,Nishimichi2009,VN2011} at $z=400$.
The initial PS in Eq.~(\ref{linear_pk}) was obtained at $z=10$ and then scaled back to the initial epoch ($z=400$) using the linear growth factor.
We used a tree-particle-mesh code, {\tt GreeM}~\cite{Ishiyama2009}, to follow the non-linear gravitational evolution. 
The gravitational softening length was set to $5 \, \%$ of the mean particle separation.
The number of particle-mesh grid cells was set to $N_{\rm p}/8$ in all the runs.

The particle position data were stored at $z=10,17,23,30,40,50,60$ and $100$.
To measure the density contrast $\delta(\bfx;z)$, we assigned the particles to the $2816^3$ grid cells in the box using the cloud-in-cell interpolation (e.g.~\cite{Jing2005,Sefusatti2016}). 
Then, the Fourier transform of $\delta(\bfx;z)$ was obtained using a fast Fourier transform\footnote{FFTW3 (the Fastest Fourier Transform in the West) at \url{http://www.fftw.org/}.}.
To explore smaller scales, we applied the folding method~\cite{Jenkins1998}, which folds the particle positions $\bfx$ into a smaller box of side length $L/n$ by replacing \rev{$\bfx$ with} $\bfx\%(L/n)$, where $a \% b$ is the reminder of $a/b$.
Here, we set $n=10$ and $100$.
This procedure effectively increases the spatial resolution by $n$ times.

The PS estimator was measured as
\beq
   \hat{P}(k;z) = \frac{1}{N_{\rm mode}} \sum_{|\bfk^\prime| \in k} \left| \, \tilde{\delta}(\bfk^\prime;z) \right|^2,
\label{pk}
\eeq
where $N_{\rm mode}$ is the number of Fourier modes in a spherical shell of $k-\Delta k/2< |\bfk^\prime| < k+\Delta k/2$.
The bin width was set to $\Delta \log_{10} k = 0.2$. 
We did not subtract the Poisson shot noise, $P_{\rm sn}=L^3/N_{\rm p}$, from the measured $P(k;z)$ because this simple formula $P_{\rm sn}$ is inaccurate, especially for high $z$ 
(see e.g. Sections 4 and 6.2 in Ref.~\cite{Heitmann2010}).

To reduce the sample variance for the HR runs, we employed the `pairing and fixing' technique~\cite{Pontzen2016,AP2016} in which paired simulations are prepared in each run. 
In the initial condition \RT{for both of the paired runs}, the amplitudes of the density contrasts in the Fourier space 
are given to reproduce the input $P_{\rm L}(k;z)$ without Gaussian randomization \RT{(i.e. $|\tilde{\delta}(\bfk)|=P_{\rm L}^{1/2}(k;z))$}. 
\RT{The phase,  $\theta(\bfk)={\rm arg}[\tilde{\delta}(\bfk)]$, for one of the paired runs is randomly chosen in a range of $0$--$2 \pi$, whereas the phase is set to $-\theta(\bfk)$ for the other run (i.e. these phases are opposite to each other).} 
Accordingly, the mean PS of the paired runs agrees with the ensemble average of many Gaussian realizations even in the non-linear regime~\cite{AP2016}. 
For the LR runs, we prepared four independent realizations with different seeds for the Gaussian initial condition.

\section{Results}

This section presents the simulation results for the non-linear PS (Subsection III.A) and the resulting boost factor (Subsection III.B). 

\subsection{Non-linear PS}

\begin{figure*}
\centering\includegraphics[width=14cm]{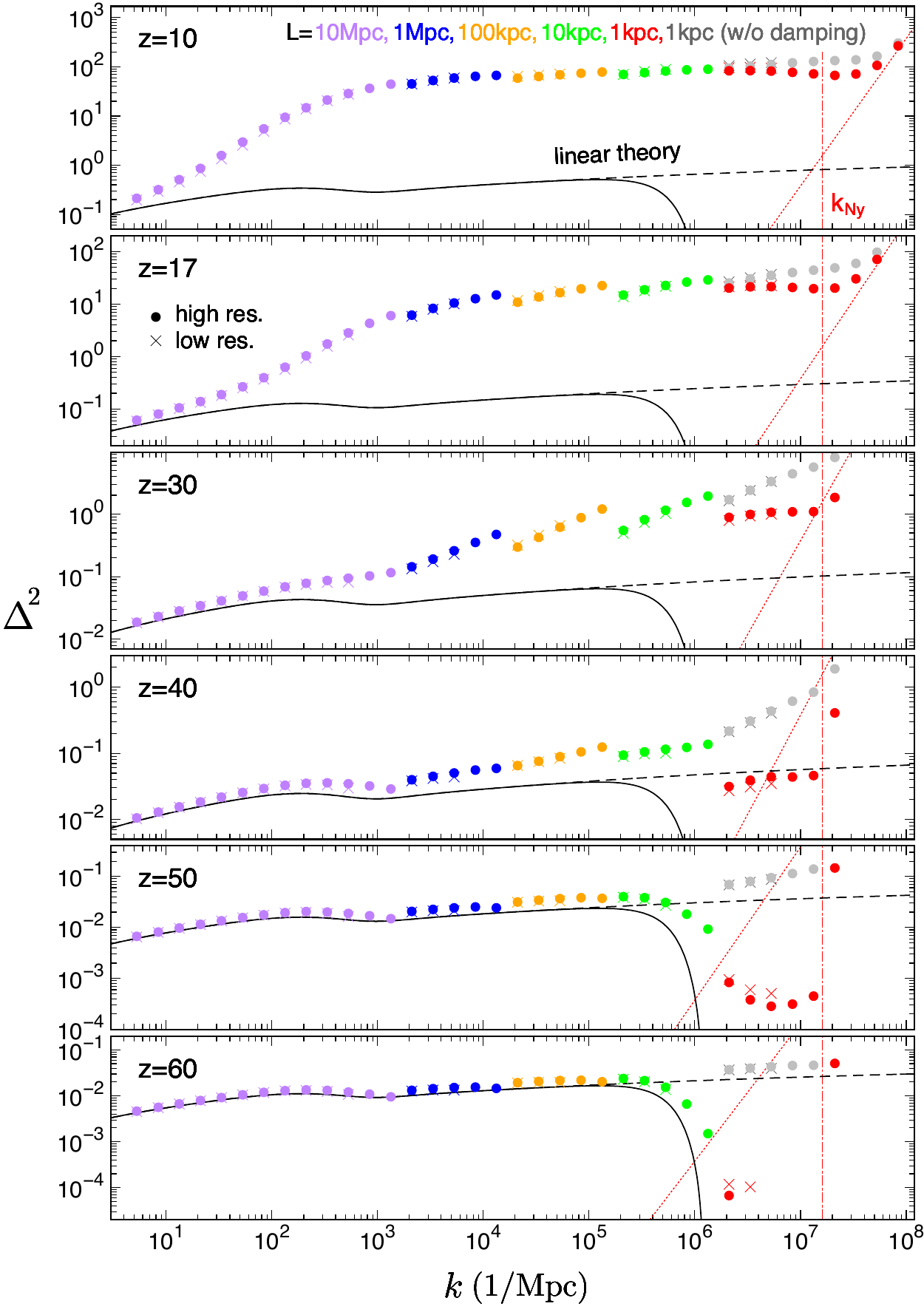}
\caption{Dimensionless matter power spectrum, $\Delta^2(k;z)$, at $z=10$--$60$. The symbols denote the simulation results with various box sizes: $L=10 \, {\rm Mpc}$ (purple), $1 \, {\rm Mpc}$ (blue), $100 \, {\rm kpc}$ (orange), $10 \, {\rm kpc}$ (green) and $1 \, {\rm kpc}$ (red) from left to right. The gray symbols are the same as the red symbols but do not include the free-streaming damping. The filled circles indicate the high-resolution (HR) results with the number of particles $N_{\rm p}=5120^3$, whereas the crosses indicate the low-resolution (LR) results with $N_{\rm p}=2560^3$. Solid curves are the linear theory prediction with the free-streaming damping, and dashed curves are the same without the free-streaming damping. Dotted red lines indicate the shot noise for the red circles. \rev{Vertical dot-dashed red lines indicate the Nyquist wavenumber for the red circles.} The discontinuity between the larger and smaller boxes, which is especially prominent at $z=17$--$40$, results from the lack of density fluctuations larger than the smaller box size.}
\label{fig_delk}
\end{figure*}

Figure~\ref{fig_delk} shows a plot of $\Delta^2(k;z)$ measured from the simulations with various box sizes ($L=1 \, {\rm kpc}$--$10 \,$Mpc), as denoted by the different colored symbols. 
Here, the results are the averages from the paired simulations (the four realizations) for the HR (LR) runs.
The plotting range is from $k=k_{\rm Ny}/10$ to $k_{\rm Ny}$, as given in Table~\ref{table_Nbody}, for the HR runs.
\rev{Only for $L=1 \,$kpc, the results are plotted up to $k=10^8 \, \Mpc^{-1}$ ($\simeq 6 \, k_{\rm Ny}$)}.
The range is the same for the LR runs, but the maximum wavenumber is the LR $k_{\rm Ny}$, which is half the HR $k_{\rm Ny}$.
Only for $L=10 \,$Mpc, the minimum wavenumber is $5.3 \, \Mpc^{-1}$, where the relative Gaussian variance of $P(k;z)$ ($\equiv \left( 2/N_{\rm mode} \right)^{1/2}$ in Eq.~(\ref{pk})) is less than $3 \, \%$.

As seen in the figure, at $z=60$, the simulation results agree fairly well with the linear theory.
At $z \simeq 40$, the non-linear evolution starts at $k \gtrsim 10^3 \, \Mpc^{-1}$.
According to previous studies (e.g.~\cite{Diemand2005,Ishiyama2014}), the first halos with Earth mass $\approx 10^{-6} \, M_\odot \, [k_{\rm fs}/(10^6 \, \Mpc^{-1})]^{-3}$ formed around this epoch. 
At $z=10$, $\Delta^2$ is approximately $100$ times larger than the linear theory at $k \gtrsim 10^3 \, \Mpc^{-1}$.
The HR and LR runs are consistent in the plotting ranges of the scales and redshifts.
\rev{It is known for the initial PS with a small-scale damping that unphysical small halos below the free-streaming scale are formed from spurious fragmentation of filaments owing to a finite mass resolution~\cite{Wang2007,Angulo2013,Schneider2013,Ishiyama2020}. 
These halos may affect the non-linear PS at $k \gtrsim k_{\rm fs}$.
However, the agreement between the HR and LR results suggests that this can be negligible up to the LR $k_{\rm Ny}$.}
The discontinuities between the larger and smaller boxes are due to the lack of density fluctuations larger than the smaller box size.
The large-scale power deficit suppresses small-scale clustering because the power transfers from large to small scales via the mode coupling between the different scales~\cite{BP1997,Padman2006,BP2009,NY2013,Nishimichi2016}.
In other words, our small box simulations give lower bounds on $\Delta^2$ (the effect of density fluctuations larger than the box size are discussed in Subsection IV.A).

The free-streaming damping at $k_{\rm fs}=10^6 \, {\rm Mpc}^{-1}$ imposed in the initial conditions persist at $z \gtrsim 50$.
However, at $z=40$ and later, this feature disappears.
For example, the results with and without initial damping, denoted by the red and the gray circles, respectively, become similar at lower $z$. 
This is because the power flow from large to small scales erases the damping feature.  
This trend is also observed in the non-linear evolution of the free-streaming damping for warm dark matter~\cite{Little1991,WC2000,SM2011,Viel2012,Inoue2015,Leo2018}.
The disappearance of the damping has important implications for $B(z)$ because the integration in Eq.~(\ref{boost_fac}) does not appear to converge in the high-$k$ limit. 

The red dotted lines in Fig.~\ref{fig_delk} indicate the shot noise, $\Delta^2_{\rm sn}=(L^3/N_{\rm p}) [k^3/(2 \pi^2)]$, for the red circles. 
The simulation results do not approach these lines \rev{at $k < k_{\rm Ny}$},
which means that the simple shot noise term, $\Delta_{\rm sn}$, is not appropriate, which is consistent with the previous remark (e.g. \cite{Heitmann2010}). 
\rev{In fact, the initial condition at $k < k_{\rm Ny}$ does not contain the shot noise.}

\rev{We comment on the realizable range of $k$ in the simulations. The initial condition includes the linear PS up to $k=k_{\rm Ny}$ but it does not include any power at $k>k_{\rm Ny}$. As time evolves, via the power transfer from large to small scales, the reliable range extends to higher $k$ ($>k_{\rm Ny}$), possibly up to the wavenumber determined by the softening length $\epsilon$ ($k_{\rm soft}=\pi/\epsilon= 20 \, k_{\rm Ny}$ in our setting).
In the halo model, the maximum reliable $k$ is determined by smallest halos resolved in the simulation~\cite{Hamana2002}; therefore the mass resolution is also important (a correspondence between the wavenumber and the halo mass is briefly discussed in Subsection IV.B.). The maximum $k$ also depends on the linear spectral index~\cite{Maleubre2021}; for a redder spectrum, the reliable $k$ extends further due to the power transfer.
In our case of Fig.~\ref{fig_delk}, the red circles approach the shot noise at $k>k_{\rm Ny}$; therefore the maximum $k$ is primarily determined by the shot noise. 
}  

Before concluding this subsection, we would like to comment on the analytical predictions of $\Delta^2$ on the basis of the stable clustering ansatz.
Let the linear PS be a single power law, $\Delta^2_{\rm L}(k;z) \propto P_{\rm L}(k;z) \, k^3 \propto k^{n_{\rm L}+3}$.
Then, the corresponding non-linear PS follows $\Delta^2(k;z) \propto k^{n+3}$ with $n+3=3(n_{\rm L}+3)/(n_{\rm L}+5)$~\cite{Peebles1980}.
In our case, in Eq.~(\ref{linear_pk}), the effective spectral index, $n_{\rm eff} +3 \equiv d\ln \Delta^2_{\rm L}(k)/d \ln k$, ranges from $-0.19$ to $0.18$ at $k=10^2$--$10^5 \, \Mpc^{-1}$.
According to the stable clustering ansatz, the non-linear spectral index, $n+3$, ranges from $-0.31$ to $0.25$, which is roughly consistent with the simulation result in the strongly non-linear regime $\Delta^2 \gtrsim 30$. 

\subsection{Cosmological boost factor}

\begin{figure}
\centering
\includegraphics[width=9.5cm]{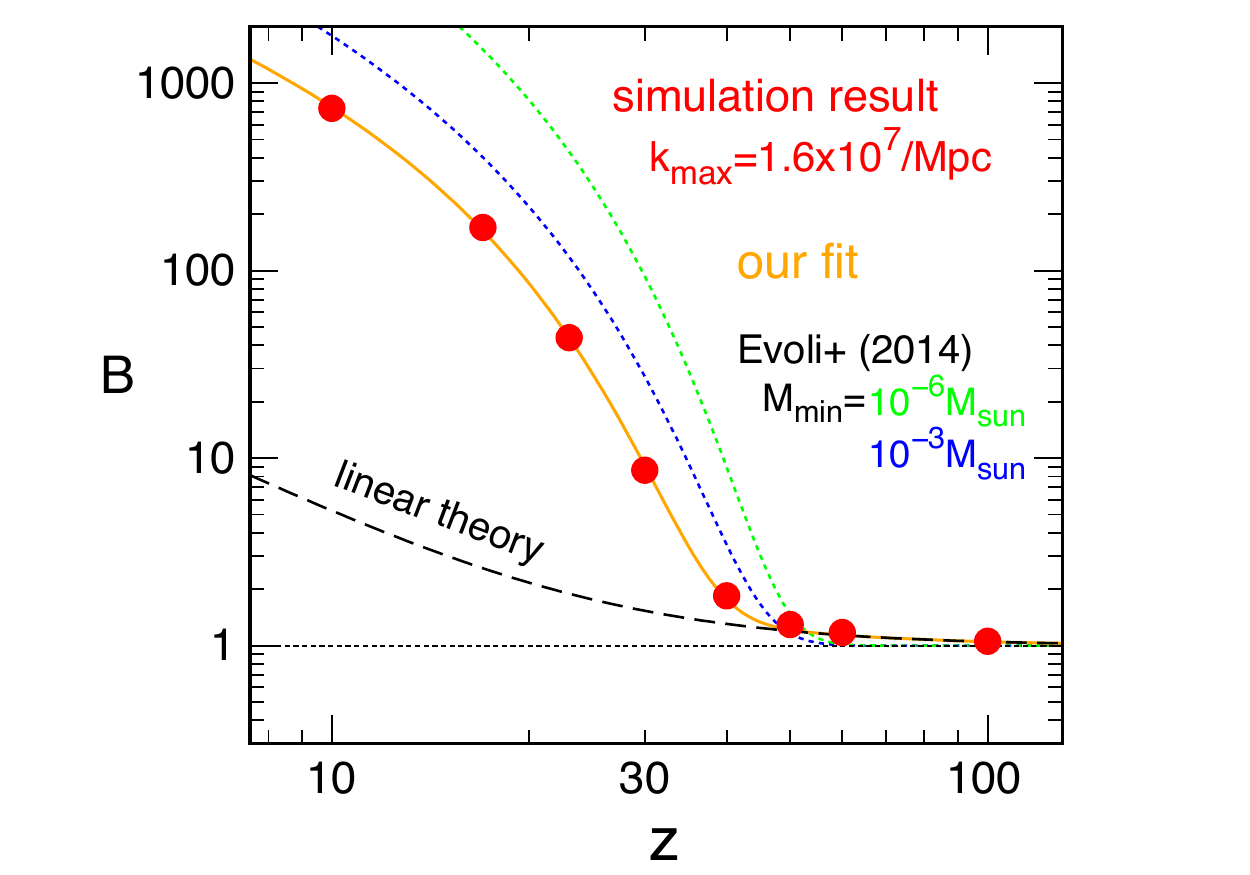}
\caption{Boost factor calculated from the simulation results of $\Delta^2$ for the maximum wavenumber $1.6 \times 10^7 \, \Mpc^{-1}$ denoted by the red circles. Orange curve represents our fit to the simulation results given in Eq.~(\ref{B_fit}), whereas dashed black curve represents the linear theory prediction. Dotted green and blue curves represent the previous halo-model results~\cite{Evoli2014} for the minimum halo masses $10^{-6} \, M_\odot$ and $10^{-3} \, M_\odot$, respectively.}
\label{fig_boost_fac}
\end{figure}

The boost factor $B(z)$ can be obtained by integrating the measured $\Delta^2(k;z)$ 
up to $k_{\rm Ny}$ ($=1.6 \times 10^7 \, {\rm Mpc}^{-1}$) for the HR run.
Here, we linearly interpolated the discrete data point of $\Delta^2(k;z)$ in Fig.~\ref{fig_delk} for the integration in Eq.~(\ref{boost_fac}).
Figure \ref{fig_boost_fac} shows a plot of the resulting $B(z)$.
The dashed curve indicates the linear theory prediction obtained analytically  from Eqs.~(\ref{boost_fac}) and (\ref{linear_pk}): $B_{\rm L}(z) = 1+514 \, (1+z)^{-2}$.
The simulation result agrees with the linear theory at $z \gtrsim 40$ but strongly increases by orders of magnitude at $z \lesssim 40$.
The orange curve represents our fit to the simulation result:
\beq
  B_{\rm fit}(z)=B_{\rm L}(z)+\frac{4.0 \times 10^4}{(1+z)^{1.27}} \, {\rm erfc} \left( \frac{1+z}{18.0} \right).
  \label{B_fit}
\eeq
In the high-$z$ limit, Eq.~(\ref{B_fit}) approaches the linear theory prediction, $B_{\rm L}(z)$.
The second term in the equation represents the non-linear correction (its functional form is the same as \rev{the one used in}~\cite{Evoli2014}, but its fitting parameters are updated).
This fitting function agrees with the simulation result within $6.8 \, \%$ at $z=10$--$100$.
Because the integration in Eq.~(\ref{boost_fac}) does not include very small-scale clustering at $k>1.6 \times 10^7 \, \Mpc^{-1}$, our $B(z)$ represents a lower bound at $z \lesssim 40$.
Our result is somewhat smaller than the previous halo-model result for $M_{\rm min}=10^{-3} \, M_\odot$ ~\cite{Evoli2014}.
Note that their result included huge uncertainties
as a result of their extrapolation of the halo properties (such as the mass function and density profile) across many orders of magnitude 
to extremely small scales. 

\rev{We include the very small-scale clustering at $k>k_{\rm Ny}$ in $B(z)$ by extrapolating the measured results of $\Delta^2$.
Suppose that $\Delta^2(k;z)$ is a single power law from $k=k_{\rm Ny}$ to a cut-off wavenumber $k_{\rm cut}$, then we have $\Delta^2(k;z)=\Delta^2(k_{\rm Ny};z) \, (k/k_{\rm Ny})^\gamma$ for $k_{\rm Ny} \leq k \leq k_{\rm cut}$ and $\Delta^2(k;z)=0$ for $k>k_{\rm cut}$. 
Figure \ref{fig_delk} suggests $\gamma \approx 0$.
Then, an additional contribution to $B(z)$, arising from $k>k_{\rm Ny}$, is written as
\begin{align}
 \Delta B_{\rm fit}(z) &= \Delta^2(k_{\rm Ny};z) \ln \left( \frac{k_{\rm cut}}{k_{\rm Ny}} \right), ~{\rm for}~\gamma=1 \nonumber \\
 &= \Delta^2(k_{\rm Ny};z) \, \frac{1}{\gamma} \left[ \left( \frac{k_{\rm cut}}{k_{\rm Ny}} \right)^\gamma -1 \right]. ~{\rm for}~\gamma \neq 1
 \label{Delta_B_fit}
\end{align}
with a fitting function
\beq
  \Delta^2(k_{\rm Ny};z) = \frac{4.1 \times 10^2}{(1+z)^{0.28}} \, {\rm erfc} \left( \frac{1+z}{16.1} \right).
  \label{delta2_kNy}
\eeq
Eq.~(\ref{delta2_kNy}) agrees with the simulation results of $\Delta^2(k_{\rm Ny};z)$ within $7.2 \, \%$ at $z=10$--$40$.
By adding $\Delta B_{\rm fit}$ to $B_{\rm fit}$ in Eq.~(\ref{B_fit}), one can obtain the boost factor for an arbitrary $k_{\rm cut}$ and $\gamma$.
For $k_{\rm cut}/k_{\rm Ny}=10,100$ and $10^3$ with $\gamma=0$, $\Delta B_{\rm fit}/B_{\rm fit}$ is less than $0.32,0.63$ and $0.94$, respectively, in the range of $z=10$--$40$; therefore $\Delta B_{\rm fit}$ does not exceed $B_{\rm fit}$ even for $k_{\rm cut}=10^3 \, k_{\rm Ny}$.
The cut-off wavenumber is currently unknown, but it can be estimated from the minimum halo mass in the halo model (see also discussion in Subsection IV.B). }

Throughout this paper, the free-streaming scale has been fixed to $k_{\rm fs}=10^6 \, \Mpc^{-1}$.
Here, we comment on the $k_{\rm fs}$ dependence on the non-linear $B(z)$.
Our simulations cover wavenumbers of up to approximately $10$ times larger than $k_{\rm fs}$ even for different $k_{\rm fs}$ values (this is determined by our simulation settings).
If the flat spectrum, $\Delta^2(k;z) \approx {\rm const.}$, continues at $k>10 \, k_{\rm fs}$, the resulting $B(z)$ would not converge and would be less sensitive to $k_{\rm fs}$. 
Additional simulations are needed to explore the $k_{\rm fs}$ dependence; however, such simulations are 
beyond the extent of this study and are left as future work.


\section{Discussion}

This section discusses the effects of density fluctuations larger than the simulation volume (Subsection IV.A), the cut-off wavenumber in the halo model (Subsection IV.B),
and the baryonic effects on $\Delta^2$ (Subsection IV.C). 

\subsection{Density fluctuations larger than the simulation volume}

\begin{figure}
\centering
\includegraphics[width=8cm]{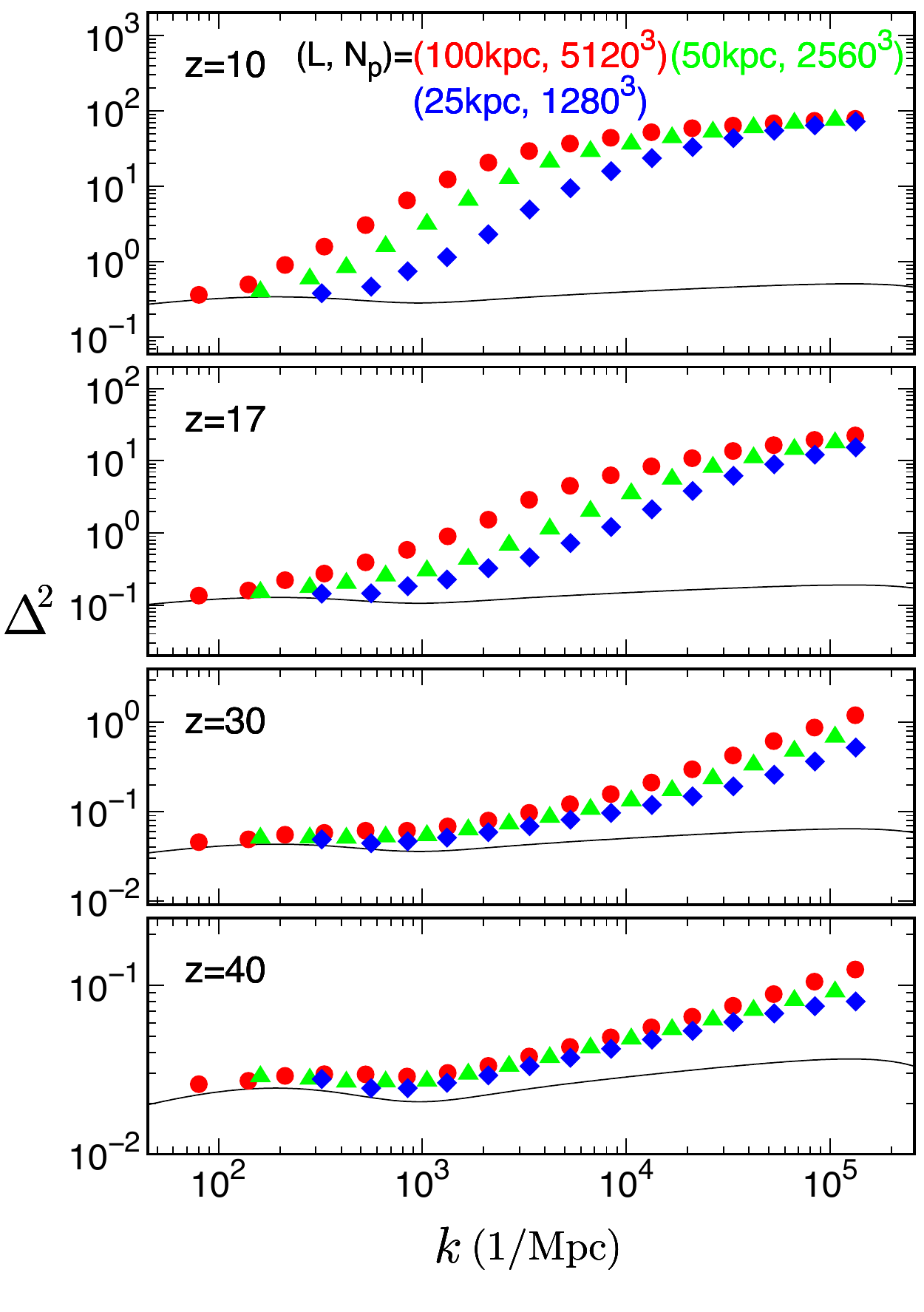}
\caption{Dimensionless matter power spectrum $\Delta^2(k;z)$ for various box sizes $L$ with different number of particles $N_{\rm p}$ but the same spatial resolution: $L=100 \, {\rm kpc}$ with $N_{\rm p}=5120^3$ (red circles), $L=50 \, {\rm kpc}$ with $N_{\rm p}=2560^3$ (green triangles) and $L=25 \, {\rm kpc}$ with $N_{\rm p}=1280^3$ (blue diamonds). Solid curves show the linear theory prediction.}
\label{fig_delk_L100k-25kpc}
\end{figure}

\begin{table}[!h]
\caption{Root-mean-square mass fluctuation $\sigma_{\rm W}$ within a cubic box of side length $L$, given in Eq. (\ref{sigma_m}), at $z=10$. Here, linear density fluctuations are assumed.}
\label{table_sigma_m}
\centering
\begin{tabular}{cccc}
\hline
$L$ & ~$\sigma_{\rm W}$ \\
\hline
$10 \, {\rm Mpc}$ & ~$0.14$  \\
$1 \, {\rm Mpc}$ & ~$0.38$  \\
$100 \, {\rm kpc}$ & ~$0.71$  \\
$10 \, {\rm kpc}$ & ~$1.09$  \\
$1 \, {\rm kpc}$ & ~$1.38$  \\
\hline
\end{tabular}
\end{table}

To examine the effects of density fluctuations larger than the box, we computed $\Delta^2$ for various box sizes while retaining the spatial resolution.
These additional simulations were run following the same procedure used in Subsection II.C.
The result is shown in Fig.~\ref{fig_delk_L100k-25kpc}.
The plotting range is from the minimum wavenumber ($=2 \pi/L$) to the particle Nyquist wavenumber.
Figure \ref{fig_delk_L100k-25kpc} indicates that the smaller box simulation underestimates $\Delta^2$ in the non-linear regime, as expected from the power flow from large to small scales. 
Here, the missing large-scale power is less important for a bluer spectrum (larger spectral index) and more important for a redder spectrum (smaller spectral index). 
Because the linear PS in Eq.~(\ref{linear_pk}) is red ($n_{\rm L} +3 \approx 0$) at $k \gtrsim 10^2 \, \Mpc^{-1}$, this effect is prominent.
Furthermore, Fig.~\ref{fig_delk_L100k-25kpc} shows that $\Delta^2$ is steep in the weak non-linear regime ($1 \lesssim \Delta^2 \lesssim 30$)  but becomes shallow in the strong non-linear regime ($\Delta^2 \gtrsim 30$).
This trend is consistent with previous findings (Subsection 5.1 in  Ref.~\cite{Smith2003}).

Next, we calculated the root-mean-square mass fluctuation $\sigma_{\rm W}$ within a cubic box in the linear theory.
This quantity $\sigma_{\rm W}$ needs to be smaller than unity to safely neglect the effect of large-scale fluctuations.
The window function for a cubic box of side length $L$ is $W(\bfx;L)$ $=L^{-3} \, \Theta(L/2-|x|)$ $\Theta(L/2-|y|)$ $\Theta(L/2-|z|)$, where $\Theta(x)$ is the step function: $\Theta (x) =1 \, (0)$ for $x \geq 0$ ($x<0$).
Its Fourier transform is $\widetilde{W}(\bfk;L)$ $={\rm sinc} (k_x L/2)$ ${\rm sinc} (k_y L/2)$ ${\rm sinc} (k_z L/2)$, where ${\rm sinc}(x)=\sin x/x$.
Accordingly, the linear mass variance can be written as
\beq
  \sigma_W^2(L;z) 
  = \int \frac{d^3 \bfk}{(2 \pi)^3} \left| \widetilde{W}(\bfk;L) \right|^2 P_{\rm L}(k;z).
\label{sigma_m}
\eeq
This variance is roughly related to $\Delta_{\rm L}^2$ as $\sigma_{\rm W}^2(L;z) \approx \Delta_{\rm L}^2 (k \! = \! 2\pi/L;z)$.
Table \ref{table_sigma_m} lists the values of $\sigma_W$ for various $L$ at $z=10$ (here, $\sigma_{\rm W} \propto 11/(1+z)$ for an arbitrary $z$).
Only for $L=10 \, \Mpc$, the missing large-scale fluctuations are safely negligible; for the other box sizes, they are not. 
For higher redshifts ($z \geq 17$), because $\sigma_{\rm W}$ is smaller, the large-scale fluctuations are less important.
Note that the linear theory, assumed in Eq.~(\ref{sigma_m}), underestimates $\Delta^2$ for $k \gtrsim 10 \, \Mpc^{-1}$ at $z=10$ and, therefore, the obtained $\sigma_{\rm W}$ indicates a lower bound. 

Density fluctuations larger than the box can be accounted for using the separate universe (SU) technique (e.g. \cite{Sirko2005,TH2013,Li2014,Wagner2015,Baldauf2016,Takahashi2019,Barreira2019,Masaki2020,Akitsu2020}).
An SU simulation can follow the non-linear clustering in an over/under-dense region of the universe.
In this technique, the mean density of the box, which is usually different from the global mean, is absorbed into the change in the cosmological parameters, i.e. the simulation runs under the `local' cosmological parameters.
For example, an over-dense region corresponds to a spatially closed universe, while an under-dense region corresponds to a spatially open universe.
The SU simulation can account for the local density contrast, as well as the external tidal field.
The SU approach, however, is beyond the scope of this paper and is left as future work.

\subsection{Cut-off wavenumber in the halo model}

This subsection estimates the cut-off wavenumber $k_{\rm cut}$ of the non-linear $\Delta^2$ from the typical size of the minimum halo (see e.g. Section 3 of Ref.~\cite{Sefusatti2014}).
The minimum halo mass $M_{\rm h,min}$ is determined by $k_{\rm fs}$ such that $M_{\rm h,min}$ $= (4 \pi/3) \, \bar{\rho} \, (\pi/k_{\rm fs})^3$ $\approx 10^{-6} \, M_\odot \, [k_{\rm fs}/(10^{6} \, \Mpc^{-1})]^{-3}$.
Because the halo is defined as a spherical region of radius $r_{\rm v}$, where the mean density is $\Delta_{\rm v}$ times higher than the background density, we have $M_{\rm h}=({4\pi r_{\rm v}^3}/{3}) \bar{\rho} \, \Delta_{\rm v}$.
Introducing the scale radius of a halo $r_{\rm s}=r_{\rm v}/c$, where $c$ is the concentration parameter, we have $k_{\rm cut} = \pi/r_{\rm s}$.
From the above equations, $k_{\rm cut}$ can be written as
\begin{align}
  k_{\rm cut} &= c \, \Delta_{\rm v}^{1/3} \, k_{\rm fs}, \nonumber \\
              &= 5.8 \, c \left( \frac{\Delta_{\rm v}}{200} \right)^{1/3} k_{\rm fs}.
\label{k_cut}
\end{align}
The typical value of $c$ is roughly $c=1$--$2$, with a large scatter comparable to its mean, for $M_{\rm h,min}$~\cite{Diemand2005,Ishiyama2014}.
Figure~\ref{fig_delk} suggests that $k_{\rm cut}$ is at least $10$ times larger than $k_{\rm fs}$.
This means, from Eq.~(\ref{k_cut}), that some halos with $c \gtrsim 2$ and/or \RT{substructure in $M_{\rm h,min}$}
would contribute to $\Delta^2$ at $k > 10 \, k_{\rm fs}$. 

In the halo model, halos with $r_{\rm s}$ primarily contribute to the non-linear $\Delta^2(k;z)$ at $k \sim 1/r_{\rm s}$.
In the small-scale limit, but larger than the cut-off scale ($k<k_{\rm cut}$), the spectral index of $\Delta^2$, $n+3$, depends on several model ingredients, including the mass function, the concentration parameter, and the linear spectral index $n_{\rm L}$ (e.g. Eq.~(4) in Ref.~\cite{MaFry2000}).\footnote{The spectral index is $n+3=[18 \beta - \alpha (n_{\rm L}+3)]/[2 (3 \beta+1)]$ where the concentration-mass relation is $c \propto M^{-\beta}$ and the mass function is  $dn/dM \propto \nu^\alpha {\rm e}^{-\nu^2/2}$, with the linear mass variance $\sigma(M) \propto \nu^{-1}$.}
If $n_{\rm L}+3=0$ and the concentration parameter is independent of the halo mass, then $n+3=0$, which is roughly consistent with our simulation result.


\subsection{Baryonic effects on $\Delta^2$}

\begin{figure*}
\includegraphics[width=8.6cm]{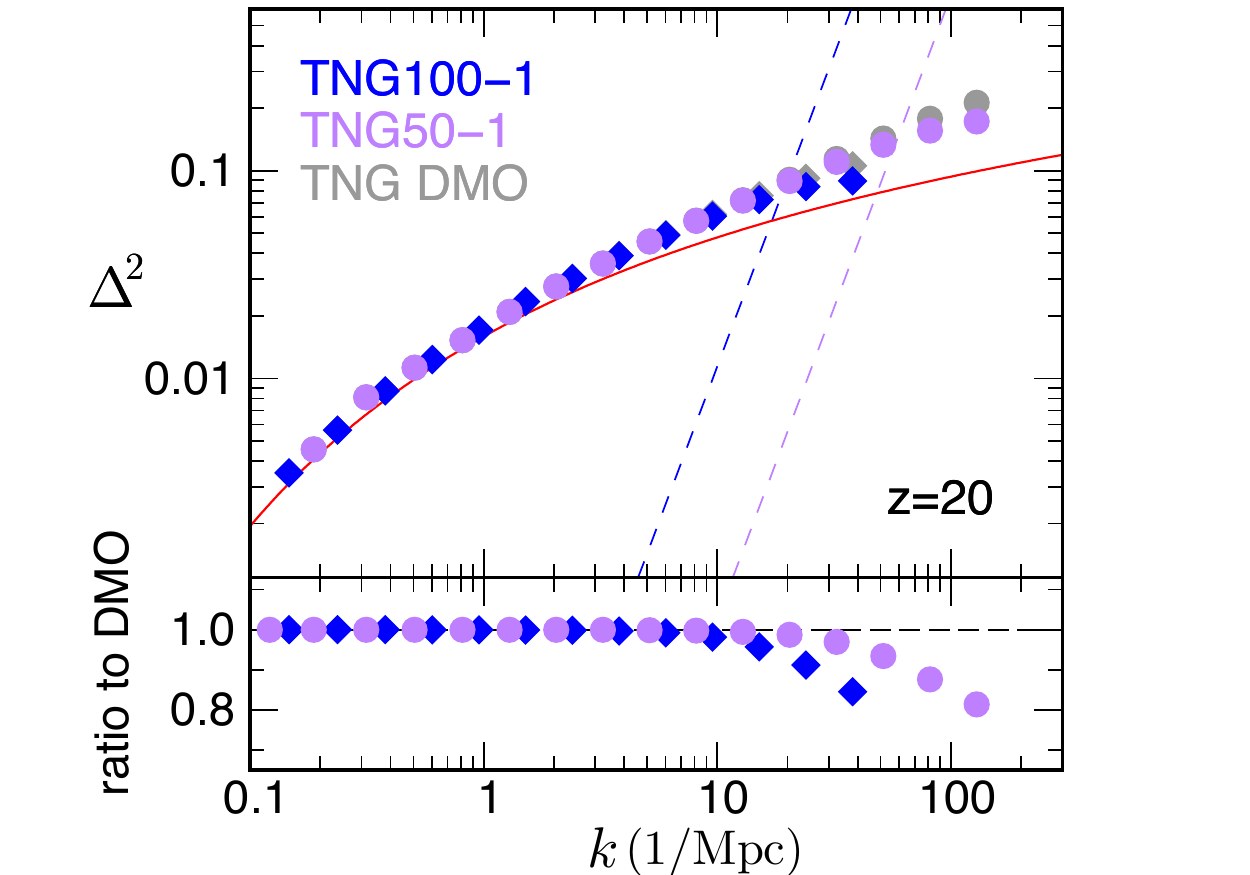}
\includegraphics[width=8.6cm]{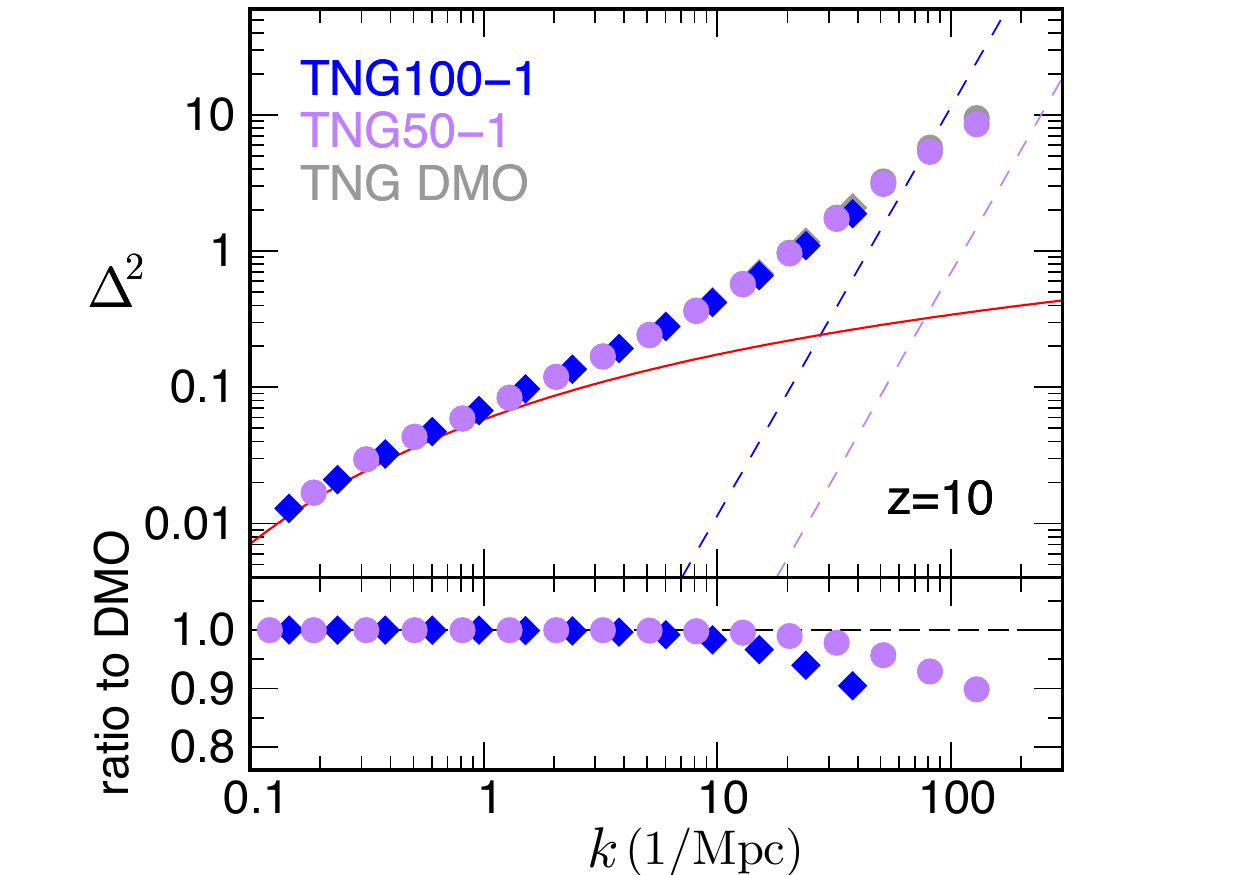}
\caption{Non-linear power spectrum measured from the hydrodynamic simulations including the baryonic processes: TNG100-1 (blue diamonds) and TNG50-1 (purple circles) at $z=20$ (left) and $10$ (right). The gray symbols are the same as the colored ones but are measured from the corresponding dark-matter-only (DMO) runs. Red curve indicates the linear theory, and dashed blue and purple lines indicate the shot noise for TNG100-1 and TNG50-1, respectively. The shot noise is included in the plotted points (i.e. it is not subtracted). The bottom portions of the panels plot the ratio of $\Delta^2$ to the corresponding value for the DMO runs.}
\label{fig_delk_TNG}
\end{figure*}

So far, we have discussed the non-linear $\Delta^2$ obtained from the dark-matter-only simulations.
However, baryonic processes (such as star formation, gas cooling and supernova and active galactic nucleus (AGN) feedback) also affect $\Delta^2$, especially at small scales (e.g. see a recent review by Ref.~\cite{Chisari2019}).
This subsection estimates the baryonic effects using public hydrodynamic simulations, IllustrisTNG\footnote{\url{https://www.tng-project.org}}.
The TNG team computed the gravitational evolution, as well as astrophysical processes, using the moving-mesh code {\tt AREPO}~\cite{Springel2010}.
They ran three sets of simulations in different size cubic boxes, with three mass resolutions for each box size.
Here, we used the highest resolution runs in the middle and small box sizes, TNG100-1~\cite{Marina2018,Springel2018,Pillep2018,Nelson2018,Naiman2018} and TNG50-1~\cite{Nelson2019,Pillep2019}, respectively.
For TNG100-1 and TNG50-1, the box sizes were $L=75 \, h^{-1} \Mpc$ and $35 \, h^{-1} \Mpc$, respectively, with the number of particles being $N_{\rm p}=1820^3$ and $2160^3$, respectively, where $N_{\rm p}$ was the same for both the baryonic and dark matter particles.
The TNG team also performed corresponding dark-matter-only (DMO) runs excluding the baryonic processes, which can be used to observe the impact of baryons on the small-scale clustering. 
Their cosmological model parameters were the same as ours.
The initial redshift was $z=127$ for all runs, and the simulation data at $z=0$--$20$ were released.
Here, we analyze the data at $z=10$ and $20$.

The upper part of the panels in Fig.~\ref{fig_delk_TNG} show plots of $\Delta^2$ calculated from the TNG simulations.
The colored symbols are from the simulations with baryons, whereas the gray symbols are from the simulations without baryons.
Here, $\Delta^2$ is calculated for the dark matter density (i.e. excluding the baryonic component) 
even in the baryonic runs because $B(z)$ is determined by the dark matter.
The potting range is up to the particle Nyquist wavenumber. 
The lower parts of the panels show the ratio of $\Delta^2$ with baryons to $\Delta^2$ without baryons.
At larger scales ($k \lesssim 10 \, \Mpc^{-1}$), as expected, the ratio is unity.
The baryons slightly suppress $\Delta^2$ by $10 \, \%$--$20 \, \%$ at $k=10$--$200 \, \Mpc^{-1}$.
\RT{The baryonic effects cannot be explored at higher $k$ ($>200 \, \Mpc^{-1}$) owing to the finite resolution of the simulations.}
It is known that at low redshifts ($z \lesssim 3$), baryons suppress the PS at $k=1$--$10 \, \Mpc^{-1}$ as a result of the AGN feedback but strongly enhance it at $k>10 \, \Mpc^{-1}$ as a result of gas cooling~\cite{Chisari2019}.
At high redshift ($z \geq 10$), because the AGN feedback is not effective, the baryon pressure should suppress small-scale clustering at $k>10 \, \Mpc^{-1}$. 

\section{Conclusions}

We obtained the cosmological boost factor, $B(z)$, at high redshifts of $z=10$--$100$ by integrating the non-linear PS measured from dedicated high-resolution $N$-body simulations.
To cover a wide range of scales ($k=1$--$10^7 \, \Mpc^{-1}$), including the free-streaming scale ($k_{\rm fs}=10^6 \, \Mpc^{-1}$), we combined five different box size simulations.
Here, our simulations cover wavenumbers up to the particle Nyquist frequency of the smallest box, $k_{\rm Ny}=1.6 \times 10^7 \, \Mpc^{-1}$.
Non-linear clustering starts at $z \simeq 40$ and enhances the PS by orders of magnitude at $z \lesssim 30$.
We found that although free-streaming damping was imposed in the initial PS, this damping feature disappears at late times ($z \lesssim 40$) as a result of the power transfer from large to small scales. 
Our $B(z)$ result agrees with the linear theory prediction at $z \gtrsim 50$ but is strongly enhanced at $z \lesssim 40$.
Our non-linear $B(z)$ is roughly consistent with, but slightly smaller than, the previous halo-model prediction with $M_{\rm h,min}=10^{-3} \, M_\odot$~\cite{Evoli2014}. 
We provide a simple fitting function for $B(z)$ in Eq.~(\ref{B_fit}).
\rev{The contribution from the small-scale fluctuations at $k>k_{\rm Ny}$ is also included in our fitting function of $B(z)$, given in Eq.~(\ref{Delta_B_fit}), using an extrapolation of the measured results $\Delta^2$ down to the smaller scales; therefore one may obtain $B(z)$ for an arbitrary cut-off wavenumber.}
Note that our $B(z)$ result is a lower bound in the non-linear epoch ($z \lesssim 40$) for the following two reasons: (i) \rev{the initial conditions of} the simulations do not include density fluctuations smaller than $2\pi/k_{\rm Ny}$ $=0.39 \, {\rm pc}$ and (ii) the lack of density fluctuations larger than the simulation volume suppresses non-linear clustering.

\section*{Acknowledgments}

We thank Tomoaki Ishiyama for his useful comments and for kindly sharing his numerical code with us.
We thank Kazuhiro Yamamoto for his useful comments on the transfer function.
We thank Nagisa Hiroshima for her careful reading and helpful comments.
Numerical computations were carried out on Cray XC50 at the Center for Computational Astrophysics, National Astronomical Observatory of Japan.
This work is supported by MEXT/JSPS KAKENHI Grant Numbers 20H05855 (RT), 20H04723 (RT), 17H01131 (RT and KK), and 20H04750 (KK).



\bibliography{refs}


\end{document}